\documentclass{iaus}
\usepackage{graphicx}

\usepackage{index}
\newindex{author}{idx}{ind}{Author Index}
\newindex{object}{odx}{ond}{Object Index}
\newindex{subject}{sdx}{snd}{Subject Index}
\makeindex

\newcommand{\aj}{\textit{AJ}}
\newcommand{\araa}{\textit{ARAA}}
\newcommand{\apj}{\textit{ApJ}} 
\newcommand{\apjl}{\textit{ApJ}}
\newcommand{\apjs}{\textit{ApJS}}
\newcommand{\aap}{\textit{A\&A}}

\newcommand{\mnras}{\textit{MNRAS}}
\newcommand{\nat}{\textit{Nature}} 
\newcommand{\pasp}{\textit{PASP}} 

\title[Disk Evolution] 
{Circumstellar Disk Evolution:  Constraining Theories of Planet Formation} 

\author[Michael R. Meyer]{Michael R. Meyer}
\index[author]{Meyer, M.R.}

\affiliation{Steward Observatory, The University of Arizona \\ 933
N. Cherry Avenue, Tucson, AZ 85721 (USA)}

\pubyear{2009}
\volume{258} 
\pagerange{109--}
\jname{The Ages of Stars}
\editors{E.E. Mamajek, D.R. Soderblom \& R.F.G. Wyse, eds.}

\begin{document}
\maketitle

\begin{abstract}Observations of circumstellar disks around stars as a
function of stellar properties such as mass, metallicity,
multiplicity, and age, provide constraints on theories concerning the
formation and evolution of planetary systems.  Utilizing ground- and
space-based data from the far--UV to the millimeter, astronomers can
assess the amount, composition, and location of circumstellar gas and
dust as a function of time.  We review primarily results from the
Spitzer Space Telescope, with reference to other ground- and
space-based observations.  Comparing these results with those from
exoplanet search techniques, theoretical models, as well as the
inferred history of our solar system, helps us to assess whether
planetary systems like our own, and the potential for life that they
represent, are common or rare in the Milky Way galaxy.
\index[subject]{solar system: formation}
\index[subject]{stars: circumstellar matter}
\index[subject]{stars: pre--main-sequence}
\index[subject]{planetary systems: protoplanetary disks}
\index[subject]{planetary systems: formation} 
\end{abstract}

\firstsection 

\section{Introduction}

Are there multitudes of planetary systems that are capable of
harboring life, like our own Solar System?  Answering this question
motivates the research activities of a great number of astronomers, as
well as scientists of many disciplines.  Yet the answer depends on
which aspect of our solar system to which one is comparing the
physical properties of other systems. Extrapolation of radial velocity
results to 20 AU suggests that planets with mass at least a third that
of Jupiter's surround 15-20 \% of sun-like stars (Cumming et
al. 2008).  Yet lower mass planets might turn out to be even more
common (e.g. Mayor et al. 2009).  Enormous progress has been made in
the past several years on many aspects of circumstellar disk evolution
(e.g. Meyer et al. 2007), especially those that can be addressed with
observations from the Spitzer Space Telescope (Werner et al. 2006).
In this review we explore answers to three key questions: 1) What is
the time available to form gas giant planets?  2) What is the history
of planetesimal collisions versus radius?  3) How do answers to the
above vary with stellar properties?  Because the answers to these
questions are subtle, one needs large stellar samples with reliable
stellar ages from the youngest pre--main sequence stars to the oldest
stars known in the galactic disk.  In our attempt to study important
evolutionary processes in the formation and evolution of planetary
systems, we assemble groups of stars with like properties (such as a
narrow range in stellar mass) as a function of age, the main topic of
this symposium.  We hope that by studying the mean (as well as the
dispersion) in those properties of the circumstellar environment as a
function of time, we can create a ''movie'' (or range of plausible
trajectories) that helps tell the story of how our solar system might
have formed.  Once completed for one range in stellar mass, we can
attempt to repeat the study for other stellar masses.  Examining the
differences in circumstellar disk evolution as a function of stellar
mass may be our best tool in delineating the most important physical
processes in planet formation.  It is a lofty goal, and often strong
assumptions are required to make progress.  We can only hope that most
of these assumptions represent hypotheses we can test in the near
future.

Observations of circumstellar gas and dust, both its amount and
geometrical distribution, can be compared to theoretical timescales
for its expected evolution.  Keplerian orbits can range from days to
millennia.  The viscous timescale in the context of an $\alpha$ disk
model depends on the orbital radius and can be $<$ 1 Myr within 10 AU
for reasonable parameters (Hartmann 1998).  Preliminary results
suggest that disk chemistry proceeds more slowly than relevant
dynamical times indicating that mixing could be important (Bergin et
al. 2007).  The inward migration of solids in the disk results in the
loss of planet--building material and remains a serious problem on
many scales: a) gas drag on meter--sized bodies can reach 1 AU/century
(Weidenschilling, 1977); b) Type I migration of lunar--mass planetary
embryos on timescales of 10$^5$ yrs; and c) Type II migration of
forming gas giant planets on timescales proportional to the viscous
time (e.g. Ida \& Lin, 2008 and references therein).  The timescale
for orderly growth of bodies through collisions (e.g. Goldreich et
al. 2004) is proportional to the product of the radius and volume
density of typical particles divided by the product of the mass
surface density of solids and the orbital frequency.  The timescale
for radiogenic heating of forming planetesimals is set by the relative
abundances of radioactive nuclides as well their half-lives (Sanders
\& Taylor, 2005).  Current models suggest that photoevaporation of gas
from illumination by EUV, FUV, and x--ray emission can disperse
primordial gas disks around sun-like stars on timescales $<$ 10 Myr
(Gorti \& Hollenbach 2009; Ercolano et
al. 2009).  Additional physical processes are important in debris disk
evolution.  In most (if not all) debris disks studied to date (our
solar system dust disk being a notable exception), mutual collisions
between dust grains reduce particle sizes to the blow--out limit where
radiation pressure efficiently removes them from the system (Wyatt 2008).  
The Lyapunov timescale characterizes divergence of orbital elements in a
chaotic system and is related to the timescale for instability in
planetary systems (Murison et al. 1994).  In this context, we note
that our own planetary system is thought to be stable on timescales
comparable to its present age of 4.56 Gyr (Laskar 1994) while the
main sequence lifetime of the Sun is approximately twice this.

In what follows, we describe some of the observational evidence for
primordial disk evolution and resulting constraints on theories of
planet formation.  We follow with a brief discussion of planet
formation, the evolution of planetesimals belts, and the dust debris
they generate.  We conclude with a summary and look forward to
exciting developments we can anticipate in the years ahead.

\section{Primordial Circumstellar Disks}

It is now well established that most sun--like stars form surrounded
by circumstellar disks.  These disks are
primordial mixtures of gas and dust initially resembling the
composition of the interstellar medium from which the star-disk
systems form.  Near--infrared excess emission traces the hottest inner
disk structures within 0.1 AU.  While the inner edge of the dust disk
is determined by the location at which the dust particles sublimate
($\sim$ 1400 K for typical silicates; Muzerolle et al. 2003), the gas
disk can extend inward, perhaps terminating near the boundary set by
magnetospheric accretion theory (e.g. Shu et al. 1994).  The
pioneering work of Strom et al. (1989) suggested that inner disks
dissipate on timescales of 10 Myr from observations of K--band excess
toward T Tauri stars in the Taurus dark cloud.  Haisch et al. (2001)
surveyed near--IR excess emission using color--color diagrams toward
hundreds of stars in several young clusters concluding that the mean
inner disk lifetime is approximately 3 Myr.  Yet if the spectral type
of star is known, and multi--color optical/infrared photometry is
available, one can carefully separate the effects of intrinsic stellar
colors, reddening due to dust along the line of sight toward the star,
and circumstellar disk excess determining with greater precision the
magnitude of any excess emission (e.g. Meyer et al. 1997).  This
approach was used by Hillenbrand and collaborators
(2008) to construct the evolutionary diagram shown
in Figure 1 which we have adapted to include: 
a) relative formation timescales for solar system
objects as determined from measurements of extinct
radioactive nuclides from meteorite samples
(Scott, 2007; Jacobsen, 2005); and
b) the frequency distribution of inner disk 
lifetimes based on these evolutionary diagrams.
While the mean disk lifetime is 3 Myr, 
there is a large {\it dispersion} of inner
disk lifetimes.  Andrews \& Williams (2005) have 
observed the distribution of disk masses
for young stars ranges over two orders of
magnitude from millimeter wave continuum data.
Perhaps the distribution of initial conditions 
(specific angular momenta in collapsing
cloud cores) results in a distribution of 
disk masses and thus disk lifetimes.

\begin{figure}[bh]
\begin{center}
 \includegraphics[width=2.4in]{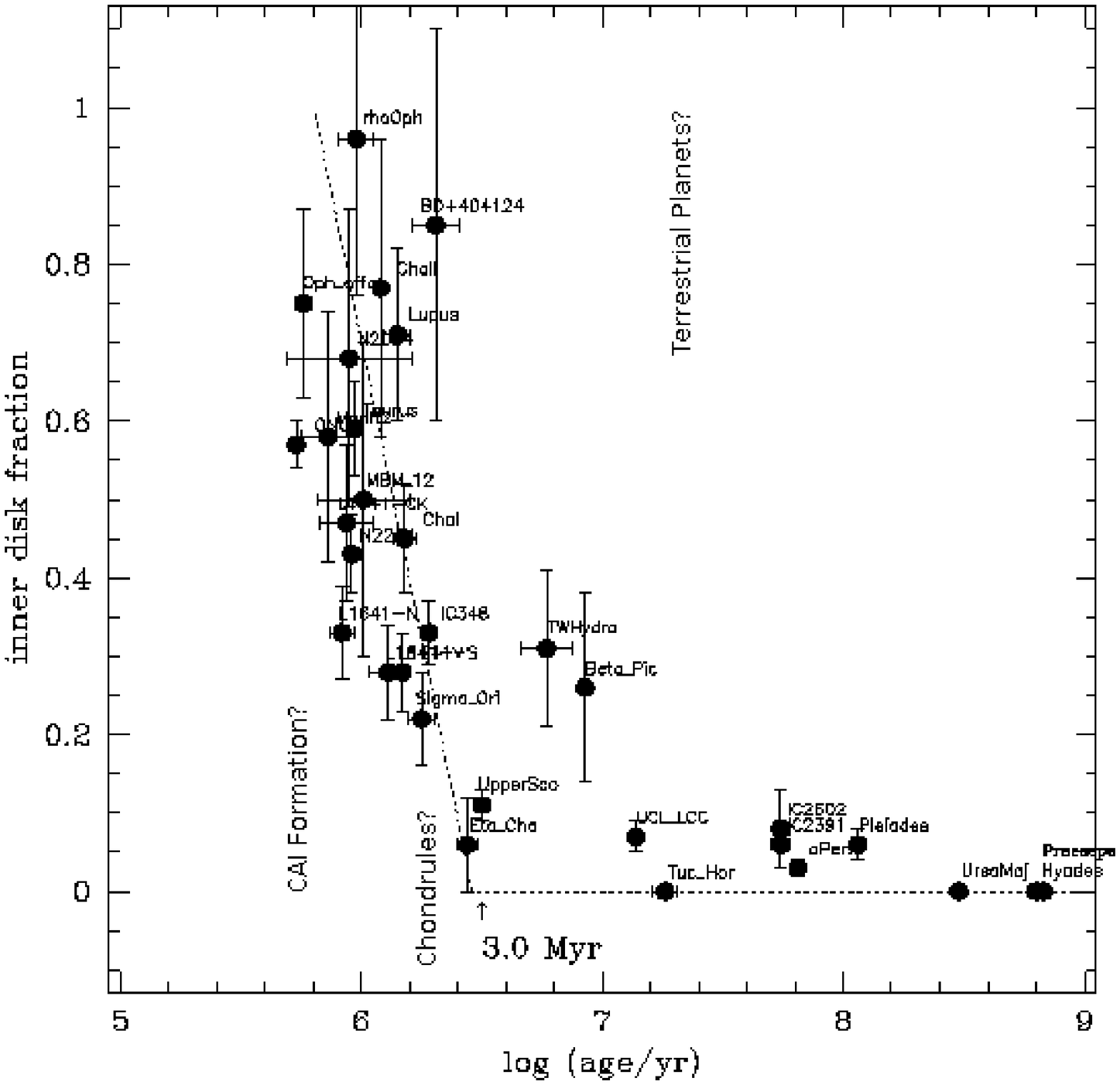}
 \includegraphics[width=2.4in]{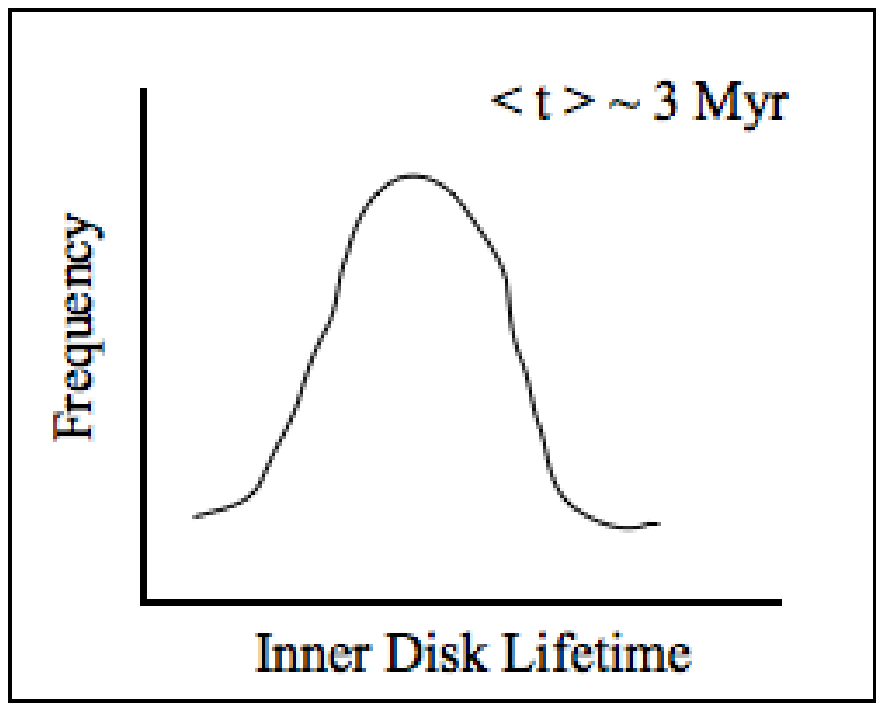}
 \caption{\textit{Left:} 
 Fraction of inner accretion disks as a function of time
  surrounding sun-like stars
 (roughly 0.5-2 M$_{\odot}$ as traced by near--IR excess emission observed
  toward stars with known spectral types.  Figure adapted from Hillenbrand
  (2008) with the approximate timescales for the formation of solar system
   objects indicated from meteoritic study of extinct radioactive nuclides.
\textit{Right:}
Schematic representation of {\it distribution} of inner disk
	  lifetimes obtained by subtracting adjacent bins in the left panel.}
\label{bigradii}
\end{center} \end{figure}

How long does it take for material in the inner disk to transition
from optically--thick to optically--thin?  Skrutskie et al. (1990)
provided some preliminary answers based on 10 $\mu$m
observations of T Tauri stars.  They
found that most stars lacking 2 $\mu$m excess also lacked
mid--infrared excess emission.  A small handful of objects, the
so--called transition objects, exhibited modest mid--IR excess but
lacked evidence for hot dust.  From the ratio of the number of objects
in transition compared to samples of T Tauri stars (few \% with
respect to T Tauri stars with primordial accretion disks, CTTS, or a
smaller fraction of young stars overall) times the typical age of the
sample stars (1--3 Myr), they estimated the duration of the transition
phase to be very short ($\sim 10^5$ years).  Subsequent work has
verified these estimates.  For example, Silverstone et al. (2006)
found no examples of inner dust disks in systems that lack signatures
of on--going gas accretion from the disk onto the star (cf. Cieza et
al. 2007).  However with surveys of star-forming regions enabled by
the Spitzer Space Telescope, dozens of these rare objects can now be
identified (e.g. Merin et al. 2008).

Much attention is focused on understanding the nature of transition
disks (Najita et al. 2007; Alexander \& Armitage 2007).  Some appear to
have inner cleared regions due to the presence of previously unseen
faint companions at small orbital radii (Ireland \& Kraus, 2008).
Work continues to understand the detailed distribution of gas and dust
in these systems (e.g. Espaillat et al. 2007).  One
troubling feature in all of these disk studies is the general {\it
lack} of correlation between key physical variables and observational
properties of the disks (Watson et al. 2009).  
For example, Pascucci et al. (2008) found no
difference in dust properties between binary and single stars.
Some disks can be quite long lived (more
than 10 Myr old) and still appear very similar to typically much
younger classical T Tauri stars (e.g. PDS 66; Cortes et al. 2009).
Often morphological ordering of disk spectra or other properties can
be quite compelling (e.g. Bouwman et al. 2008).  But more often than
not, ranking by estimates of stellar age appear to disturb the
apparent evolutionary sequences.  Although errors in determining
stellar ages are a likely culprit, it does seem that there are
''hidden variables'' contributing significantly to primordial disk
evolution.

As a complement to studies of the dust, can we probe further the
evolution of the gas from which giant planets might form (Najita et
al. 2007)?  For young stars in the Taurus dark cloud, there is an
excellent correspondence between spectroscopic signatures of gas
accretion from the inner disk onto the surface of the star and the
presence of near--IR excess emission (Hartigan, Edwards, and Ghandour,
1995).  Indeed the evolution of accretion rates appears to mimic to
some extent the evolution of hot dust (Hartmann et al. 1998; Gatti et
al. 2008).  Spitzer spectra are starting to
reveal interesting chemistry in the disks young young stars (Carr \&
Najita, 2008; Pascucci et al. 2009).  High resolution near--infrared
spectroscopy combined with high spatial resolution can reveal
interesting offsets between the emitting regions of various molecules
(Pontoppidan et al. 2008).  Millimeter wave observations are required
to trace rotational transitions of cool gas at large radii (e.g. Dutrey
et al. 2007).

Yet the bulk of the mass in these disks is in molecular hydrogen,
which is difficult to detect.  Some efforts have been made to trace
molecular hydrogen directly by observing the pure rotational lines in
the mid--infrared from the ground at high spectral resolution (Bitner 
et al. 2008).  The Spitzer Space Telescope also provided a platform to
search for gas using the IRS in high resolution mode.  Based on
equilibrium chemical models of Gorti \& Hollenbach, 2004), Pascucci et
al. (2006) report non--detections from the FEPS survey (Meyer et
al. 2006) for stars with ages between 3--100 Myr that lack signatures
of accretion, but possess optically--thin dust emission.  The upper
limits place constraints of $<$ 10
Myr on the timescale to form gas giants in these systems.  These
timescales are of interest for comparison to models of gas giant
planet formation through classical core accretion (e.g. Lissauer \&
Stevenson 2007) as well as gravitational fragmentation which proceeds
more quickly (Durisen et al.  2007).  Future observations with
Herschel and SOFIA will be powerful
probes of even small amounts of residual emission perhaps placing
constraints on the formation of super-earths and ice giants.  We note
for completeness that circumstellar gas has been detected for some
debris disks (e.g. Dent et al. 2005), which are
the subject of the next section.

\section{Planet Formation and Generation of Debris Dust}

How does remnant disk material evolve when the bulk of the gas capable
of forming giant planets has dissipated?  In the classical theory,
protoplanets initially grow in an orderly way through collisions of
comparable sized bodies.  As gravitational focusing becomes important,
the larger bodies grow fastest in a runaway mode until a modest number
of ''successes'' reach the local isolation mass, essentially consuming
all material within several times their gravitational sphere of
influence (Hill radius).  These oligarchs then perturb each other over
time resulting in handfuls of collisions that create a
small number of planets 
(e.g. Nagasawa et al. 2007).  At 1 AU around the Sun, we imagine the
Earth as built through the collision of several Mars-sized 
objects.\footnote{The last of these collisions resulting in the
formation of the Moon (e.g. Canup 2004).} It is thought that
similar processes operate (in the presence of gas) to form the cores
of giant planets. In the first phase, we can estimate the frequency of
collisions by knowing the surface density of solids ($\sigma$) as well
as the orbital timescale ($\Omega$):
$$ \tau \sim (R_{body} \times \rho_{body})/(\sigma \times \Omega)$$
where R and $\rho$ are the radius and volume density of the object
built through these collisions in time $\tau$.  If we combine estimates
of the dependence of disk mass on star mass ($\sigma_{disk} \sim M_*$)
and orbital radius ($\sigma_{disk} \sim 1/a$) with the orbital frequency 
we get $ \tau \sim a^{5/2} M_{*}^{-3/2}$
for a constant $\rho$.  This implies, within a fixed time interval,
disks around stars of higher mass form:
a) more massive planets at a given orbital distance; and b) planets of a given mass 
at larger orbital separation.  Both are consistent with recent radial 
velocity results (Johnson et al. 2007).   Yet it appears that primordial 
disks evolve more quickly around stars of higher mass 
(Carpenter et al. 2006), complicating the implications of this simple
picture for planet formation around stars as a function of their mass. 

Observations of debris disks have become powerful tools to study the
evolution of planetary systems and their value has increased as we
have learned more about the asteroid and Kuiper belts in our own Solar
System (Wyatt 2008).  As cooler dust orbiting at larger radii emits
at longer wavelengths, we can use the wavelength dependence of excess
emission as a proxy for its location in the absence of resolved
images.  With the Spitzer Space Telescope, we can detect $>10^{-5}$
Earth masses of dust if it is found in micron--sized particles from
photometric surveys at 24 $\mu$m, efficiently tracing material with
temperatures above 100 K (often tracing radii $<$10 AU around
sun--like stars).  Meyer et al. (2008) report the results from FEPS
for 24 $\mu$m excess emission around an unbiased parent sample of 309
stars.  They find that 10--20 \% of sun--like stars show evidence for
24 $\mu$m excess over an age range from 3--300 Myr with a significant
drop in the frequency of 24 $\mu$m excess for older stars.  As we can
only observe the product of the frequency of this phenomena and its
duration, the interpretation of these results is ambiguous.  
They also suggest that the frequency of excess
{\it may} be higher for stars in open clusters compared to field stars
of comparable age though the evidence for this is not significant.
Note that the ages of the open cluster stars are the ''gold standard''
for stellar ages in these studies and thus it is difficult to compare
results between them and field stars.  In contrast, Greaves et
al. (2009) find a surprising lack of emission at 1.3 mm towards
sun--like stars in the Pleiades open cluster despite significant 
improvement in sensitivity compared to previous observations.

The above mentioned Spitzer results are consistent with other surveys
for 24 $\mu$m excess around FGK stars as a function of time
(e.g. Siegler et al. 2007).  It is worth
remembering that these survey limits are only able to detect dust
producing planetesimal belts that are about $\times 1000$ brighter
than our own inner zodiacal dust generated from collisions in the
asteroid belt.  Wyatt et al. (2007) compare these observations to
models of collisional evolution for planetesimal belts finding that
some warm debris excesses (e.g. HD 69830) are so large that they
cannot be the result of pure collisional evolution and must
be transient.  Although the models of Kenyon \& Bromley (2004) explain the
observed behavior in a qualitative way, 
updates to the input physics continue to
improve the agreement between the models and data (Bromley \& Kenyon
2008).  Models for several debris disks detected with Spitzer require
extended distributions of dust 
(Hillenbrand et al. 2008) consistent with resolved images of some
sources (Corder et al. 2009).

What (if any) correlation should we expect between the presence of
dust debris and the frequency of gas giant planets detected through
radial velocity variations?  One might speculate that a disk rich in
heavy elements capable of forming gas giant planets (e.g. Fischer \&
Valenti 2005; Santos et al. 2004) might initially have a high
surface density of solids and thus produce a lot of dust early--on,
outshining a system that lacked enough dust to form giant planets.  If
however, such a system produces several planets unstable
to mutual perturbations, dynamical rearrangement could deplete
dust-producing planetesimals in the inner and outer regions 
as suggested by the Nice model for the
early evolution of our Solar system (e.g. Tsiganis et al. 2005).  Such a
system could then become a very weak dust producer at late times
(Figure 2).  It is in part these complexities that prevent us from
making comparisons of Spitzer observations
to what is known about our Solar system.

\begin{figure}[htb]
\begin{center}
\includegraphics[angle=0, width=\textwidth]{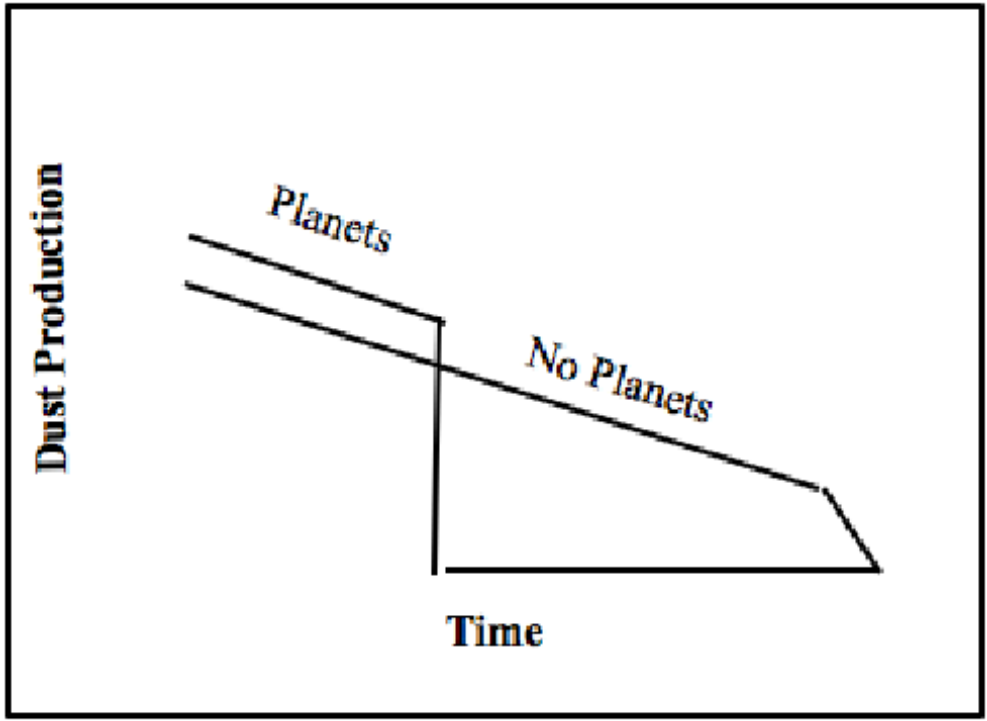} 
 \caption{ Graphical representation of dust production versus time for 
  the case of a disk with an initially high surface density of solids that
  forms gas giant planets and another with a lower initial surface density
  that does not.  The planet-forming disk outshines the other disk early
  on, but then suffers a dynamical event that removes most 
  of the dust-generating planetesimals.  The disk that was never able to 
  form giant planets might generate more dust at late times.}
   \label{fig:f2}
\end{center}
\end{figure}

Moro--Martin et al. (2007a) search for a correlation between the
presence or absence of excess with gas giant planets and was unable to
confirm any (though the sample size was modest).  Apai et al. (2008)
also searched (without success) for the presence of gas giant planets around
stars for which large inner holes in the dust debris had been
inferred.  There are, however, remarkable examples of systems with
evidence for planets that also maintain debris which deserve special
study (Moro--Martin et al. 2007b; Lovis et al. 2006) including the
recently announced direct imaging results for Fomalhaut (Kalas et
al. 2008) and HR 8799 (Marois et al. 2008).  We can also ask whether
debris is correlated with the heavy element abundance of the star as
it is famously with the probability of having a gas giant planet.
Greaves, Fischer, \& Wyatt (2006) find that the metallicity
distribution of stars with cold debris detected in the sub--millimeter
is consistent with having been drawn from the same distribution as the
parent sample of stars.  We find a similar result for stars in the
FEPS sample (Najita et al., in preparation).  Perhaps gas giant planet
formation is a {\it threshold} phenomena with regard to the heavy
element content of the primordial circumstellar disk whereas the
requirements for ending up with a dust producing debris disk are more
forgiving.  Bryden et al. (2006) estimate that Spitzer only detects
the tip of the iceberg in terms of dust debris, and survey results to
date are {\it consistent} with our Solar System IR emission being near
the mean for sun--like stars in the Milky Way.

Given the mixed results for the correlation of debris with planets,
can we expect more order in the behavior of debris with stellar
multiplicity?  There is considerable literature on the similarities
and differences in disk evolution between single and binary T Tauri
stars (Monin et al. 2007).  With regard to
debris around one or both components of a binary system, Trilling et
al. (2007) report Spitzer results finding a complex mix of behaviors
including circumbinary debris disks, circumstellar disks in wide
binary systems, as well as the apparition of dust at inferred orbital
distances consistent with the separation of the binary companion!  At
minimum, one can infer that debris dust as detected by Spitzer is not
inhibited in multiple systems.

The ''last word'' from the FEPS program comes from Carpenter et
al. (2008; 2009) in which the extant database and synthesis of results
are described respectively.  With the addition of the IRS spectra from
5--33 $\mu$m for the sample, debris suspected from the 24 $\mu$m
photometry can be confirmed with confidence.  Even more important, the
temperature of the hottest dust detected can be estimated from the
broadband spectrophotometry providing estimates of the inner radius of
the debris disks in the sample.  The detected dust is cooler on
average (and thus located at larger radii) than assumed in Meyer et
al. (2008): the distribution of inner radii for the (often extended)
debris disks detected in the survey range from 3--30 AU, with a peak
at 10 AU.  There is no evidence for evolution of the dust temperature
with age, though the magnitude of the observed excess does decline
over time (with considerable dispersion at any one age).  It appears
that some of the 24 $\mu$m excess emission observed is the Wien tail
of bright, but cool, dust at radii beyond the terrestrial planet zone.
Emission evolves from bright to faint (or in some cases hot to cool)
within 300 Myr.  Yet the picture that emerges overall is one where
most debris is cleared out to radii of 10 AU on timescales of 3--10
Myr!  Whether this represents complete ''mission success'' for planet
formation between 0.1--10 AU or presents a challenge to current theory
remains to be explored.  Constraining the outer disk radii will
require observations at longer wavelengths utilizing Herschel, SOFIA,
or more sensitive sub--millimeter observations (e.g. Roccatagliati et
al. 2009).

It is extremely tempting to compare these results for FKG stars to
those for stars of different temperature \& luminosity in order to
explore diversity in the formation and evolution of planetary systems
as a function of stellar mass.  However, such comparisons require
caution.  For example, in a survey where the sensitivity is limited by
photometric precision of detecting an excess in contrast to the flux
of the star, the amount of detectable dust around a more luminous star
is larger than the detectable amount around one of lower luminosity.
Also dust at a given temperature is located at a larger radius around
a more luminous star.  Any comparison should take this into account in
terms of solid angle for the emission as well as the expected orbital
period and dust surface density distribution.  Overall, one can say
that the magnitude of the excesses observed at 24 and 70 $\mu$m around
A stars are larger, and more common, than those observed around G
stars (Su et al. 2006), yet the timescale for
the duration of the phenomena is slightly shorter around A stars
compared to G stars (Currie et al. 2007; cf. Trilling et al. 2008). 
Gautier et al. (2007) find that debris emission around M dwarfs is
even weaker. 

\section{Summary and Future Work}

So are planetary systems like our own are common or rare among sun-like
stars in the Milky Way galaxy?  Unfortunately, we cannot yet answer
this question, but we can provide some important constraints that are
stepping--stones to future progress.
{\it Primordial Disk Evolution:} Disks around lower mass stars are
less massive and live longer than their more massive counterparts.
The large observed dispersion in evolutionary times could indicate a
dispersion in initial conditions.  Overall, disk evolution appears to
proceed from inside--out as expected.  
{\it Change you can believe in:} The duration of the transition time
from primordial to debris is $\sim$ 10$^5$ yrs.  Planetesimal belts
evolve quickly out to 3--30 AU.  Evidence for 24 $\mu$m excess is largely
gone by 300 Myr.  There is a hint that such excesses might be more
common in open clusters at a given age though more work on this is
needed.  
{\it Debris Disk Evolution:} Currently detectable extra--solar debris
systems are all collision-dominated in their evolution.  Debris is
brighter and more common around stars of higher mass.  Evolutionary
paths are diverse but the observed distributions are consistent with
our Solar System debris disk being common among stars $>$ 1 Gyr old.
The connection between debris and planetary systems is unclear.  Yet
it may turn out that debris (and perhaps terrestrial planets) are more
common than their gas giant counterparts.  As we know that nearly all
young stars begin their lives surrounded by primordial disks of gas
and dust capable of making some sort of planetary system, one wonders:
are systems without debris those with dynamically full planetary
systems, or those without any planets whatsoever?

Anyone who has taken a turn leading the blind around the proverbial
elephant (i.e. modelling spectral energy distributions) knows that a
resolved image is worth more than 1024 $\times$ 1024 photometric
points on an SED.  Constraining model parameters with resolved
emission at one or more wavelengths is vital to making progress.  In
addition to high contrast imaging of disks in scattered light from space
and ground--based telescopes equipped with adaptive optics, millimeter
wave interferometry will continue to make crucial contributions.  The
soon to be launched ESA Herschel Space Telescope will also build on
the work of Spitzer, providing dozens of newly resolved debris disks
in thermal emission.  Finally, we look forward to science observations
with ALMA and JWST that will expand our understanding of the formation
and evolution of planetary systems in ways we can scarcely imagine
(provided of course that we first solve the thorny problem of
obtaining accurate stellar ages and uncover the nearest, youngest,
sun--like stars as prime targets of observation).

\acknowledgment The author would like to thank many colleagues whose
work has contributed to his current understanding regarding the
formation and evolution of planetary systems including members of 
the FEPS/c2D/Glimpse Legacy Science Teams, the MIPS/IRS/IRAC
Instrument Teams, and the Spitzer Science Center staff.  MRM acknowledges
support from the Legacy Science Program through a contract from
NASA/JPL as well as the LAPLACE node of the NASA Astrobiology
Institute.

\begin{discussion}

\discuss{M. Liu}{Open clusters and young moving groups represent the
best ``benchmark'' systems for understanding evolution, since they are
coeval and formed in the same environment.  However, even in these
very simple samples, debris disk properties are diverse and not easily
explained.  Do you find this to be a discouraging result in attempting
to develop a comprehensive picture of debris disk evolution?}

\discuss{M. Meyer}{Evidence from debris disk studies as well
observations of exoplanets suggest that the outcomes of the planet
formation process are diverse, as are the paths of subsequent
evolution.  By combining detailed studies of individual objects (those
systems that are resolved in scattered light and/or thermal emission)
as well as large surveys conducted with ground- and space- based
telescopes (Spitzer, Herschel and JWST) we can hopefully discern the
overall climates that are conducive to planet formation, in contrast
to our inability to predict the prospects of planet formation for any
system (the ``weather'').}

\discuss{R. Jeffries}{Could the low frequency of debris disks be
explained by ``episodic'' dust production?  In other words, could
debris disks be more common but only episodically produced?}

\discuss{M. Meyer}{It is true that we only observe the product of the
frequency of the debris disk phenomenon and the distribution of
durations.  Indeed, if the warm debris disk epoch lasts a short time,
the overall frequency could be much higher than observed (Meyer et al
2008).  However observational support for this scenario is lacking
(Carpenter et al. 2009).  Wyatt et al. (2007) explore conditions
under which an observed debris disk can be unambiguously identified as
transient.}

\end{discussion}
\end{document}